\documentclass[a4paper, 12pt]{article}
\usepackage{graphicx}
\def \d {{\rm d}}

\begin{document}

\title{\bf The $C$-metric as a colliding plane wave space-time}

\author{J. B. Griffiths and R. G. Halburd
\\ \\ \small
Department of Mathematical Sciences, Loughborough University, \\
\small Loughborough,  Leics. LE11 3TU, U.K. 
\\ \\
\small E-mail: J.B.Griffiths@lboro.ac.uk, R.G.Halburd@lboro.ac.uk}

\date{\today}
\maketitle

\begin{abstract}
\noindent
It is explicitly shown that part of the $C$-metric space-time inside the black hole horizon may be interpreted as the interaction region of two colliding plane waves with aligned linear polarization, provided the rotational coordinate is replaced by a linear one. This is a one-parameter generalization of the degenerate Ferrari--Ib\'a\~nez solution in which the focussing singularity is a Cauchy horizon rather than a curvature singularity. 

\end{abstract}

\section{Introduction}

In 1986, Chandrasekhar and Xanthopoulos~\cite{ChaXan86b} showed that part of the Kerr space-time in the time-dependent region between the two horizons can also be interpreted as representing the interaction region of two colliding plane waves when the periodic rotational coordinate is replaced by an infinite linear one. In this case, the singularity that is caused by the mutual focussing of the two waves, and is generically a (scalar polynomial) curvature singularity, is replaced by a Cauchy horizon that corresponds to the inner (Cauchy) horizon of the Kerr space-time. An alternative region of this space-time which can have the same interpretation, but in which the focussing singularity corresponds to the outer (event) horizon, was pointed out by Hoenselaers and Ernst~\cite{HoeErn90}. These authors considered extensions of the space-time through the horizon either towards the ring singularity or towards the asymptotically flat region respectively, but such extensions are not uniquely determined. (For a review of colliding plane wave space-times and their properties, see~\cite{Griff91}.)

The Schwarzschild limits of these colliding plane wave space-times have been analysed by Ferrari and Ib\'a\~nez~\cite{FerIba87b, FerIba88}, who also extended them in the equivalent interpretation of parts of the Taub--NUT space-time in the time-dependent (Taub) region. With the addition of a charge parameter, Chandrasekhar and Xanthopoulos~\cite{ChaXan87b} and Papacostas and Xanthopoulos~\cite{PanXan89} have further shown that parts of the Kerr--Newman and Kerr--Newman--NUT solutions respectively also have similar interpretations. It is therefore natural to ask whether or not the $C$-metric can also be interpreted in this way. This is a different one-parameter type~D generalization of the Schwarzschild solution, which admits the same symmetries and the same internal black hole structure. The purpose of the present article is to demonstrate explicitly that the equivalent interpretation of the $C$-metric is possible for an appropriate region, and to display its properties as a colliding plane wave space-time.

\section{The $C$-metric} 

The solution that is known as the $C$-metric was originally found in its static form by Weyl in 1917~\cite{Weyl17} and subsequently rediscovered many times. Its basic properties have been interpreted by Kinnersley and Walker~\cite{KinWal70} and Bonnor~\cite{Bonnor83}, who showed that its analytic extension represents a pair of black holes which accelerate away from each other due to the presence of strings or struts that are represented by conical singularities along the axis of symmetry. This solution is characterised by a certain cubic function. Recently, Hong and Teo \cite{HongTeo03} presented a new parametrization of this function which simplifies its root structure. This has been shown \cite{GrKrPo06} to help both with calculations and with the physical interpretation of this space-time.

Using the more transparent form given in \cite{GrKrPo06}, the $C$-metric can be expressed as 
 \begin{equation} 
 \d s^2={1\over(1+\alpha r\cos\theta)^2} \left( -Q\,\d t^2 
 +{\d r^2\over Q} +{r^2\,\d\theta^2\over P} +P\,r^2\sin^2\theta\,\d\varphi^2 \right), 
 \label{Cmetric} 
 \end{equation}  
 where 
 \begin{equation} 
  P=1+2\alpha m\cos\theta, \qquad 
  Q=(1-\alpha^2r^2)\Big(1-{2m\over r}\Big), 
 \label{BLPQ} 
 \end{equation} 
 $0<2\alpha m<1$, \ and \ $\varphi\in(-\pi C,\pi C)$, \ where $C$ is a constant that determines the distribution of the topological singularities along the axis of symmetry. This family of solutions reduces precisely to the familiar form of the Schwarzschild solution when ${C=1}$ and ${\alpha=0}$.

In order to interpret part of this space-time as representing the interaction region of two colliding plane waves, it is appropriate to make the initial coordinate transformation 
 \begin{equation} 
 r=m(1+\eta), \qquad \cos\theta=\mu, \qquad t=x, \qquad \varphi={y\over m}, 
 \label{firsttrans} 
 \end{equation} 
 where we now take $x,y\in(-\infty,\infty)$. The time-dependent region inside the black hole is thus given by ${|\eta|<1}$, in which the limit ${\eta=1}$ corresponds to the event horizon. With (\ref{firsttrans}), the metric (\ref{Cmetric}) takes the form 
 \begin{equation} 
 \begin{array}{l}
 {\displaystyle \d s^2 ={m^2(1+\eta)^2\over\big[1+\alpha m\mu(1+\eta)\big]^2} \left[ 
 -{\d\eta^2\over(1-\eta^2)\Big[1-\alpha^2m^2(1+\eta)^2\Big]}
 +{\d\mu^2\over(1-\mu^2)(1+2\alpha m\mu)} \right]} \\[20pt]
 \hskip4pc +\> e^{-U}\Big(e^V\,\d x^2+e^{-V}\,\d y^2\Big), 
 \end{array} 
 \label{etamuMetric} 
 \end{equation} 
 where 
 $$ \begin{array}{l}
 {\displaystyle e^{-U}= { \sqrt{1-\eta^2} \sqrt{1-\mu^2} 
 \sqrt{1-\alpha^2m^2(1+\eta)^2} \sqrt{1+2\alpha m\mu}
 \over \big[1+\alpha m\mu(1+\eta)\big]^2}, } \\[16pt]
 {\displaystyle \ e^V= { \sqrt{1-\eta} \sqrt{1-\alpha^2m^2(1+\eta)^2} 
 \over (1+\eta)^{3/2} \sqrt{1-\mu^2} \sqrt{1+2\alpha m\mu}}. } 
 \end{array} $$

To facilitate the introduction of appropriate double null coordinates, it is first convenient to introduce the parameter $k$ such that \ ${\alpha m=k/(1+k^2)}$ \ where ${0<k<1}$. \ Convenient timelike and spacelike coordinates $T$ and $Z$ can then be introduced by putting 
 $$ \eta={s_1+k^2\over1+k^2s_1}, \qquad \mu={s_2-k\over1-k\,s_2}, $$ 
 where $s_1$ and $s_2$ are the Jacobi elliptic functions 
 $$ s_1={\rm sn}(T-T_0,k), \qquad
 s_2={\rm sn}(Z+Z_0,k), $$ 
 and $T_0$ and $Z_0$ are constants whose values must be chosen appropriately\footnote{
      It is possible, for example, to chose $T_0$ and $Z_0$ such that \ ${{\rm sn}(T_0,k)=k^2}$ \ and \ ${{\rm sn}(Z_0,k)=k}$. \ With this choice \ ${T=0}$, ${Z=0}$ \ would corresponds to \ ${\eta=0}$, ${\mu=0}$, \ but this turns out not to be convenient. }.
      With this 
 $$ {\d\eta\over\d T}=\sqrt{1+k^2}\sqrt{1-\eta^2}\sqrt{1-\alpha^2m^2(1+\eta)^2}, \qquad
 {\d\mu\over\d Z}=\sqrt{1+k^2}\sqrt{1-\mu^2}\sqrt{1+2\alpha m\mu} , $$ 
 and the metric (\ref{etamuMetric}) becomes 
 \begin{equation} 
 \begin{array}{l}
 {\displaystyle \d s^2={1+k^2\over(1+ks_1s_2)^2}\Bigg\{ 
 \left[ m\left({1+k^2\over1-k^2}\right)(1+s_1)(1-ks_2) \right]^2 \Big(-\d T^2+\d Z^2\Big) } \\[12pt]
 \hskip10pc {\displaystyle +{{c_1}^2{d_1}^2(1-ks_2)^2\over(1+k^2)^2(1+s_1)^2}\,\d x^2 
 +{(1+s_1)^2{c_2}^2{d_2}^2\over(1-ks_2)^2}\,\d y^2 \Bigg\}, } 
 \end{array}
 \label{TZMetric} 
 \end{equation} 
 where ${c_1={\rm cn}(T-T_0,k)}$, ${d_1={\rm dn}(T-T_0,k)}$, ${c_2={\rm cn}(Z+Z_0,k)}$ and ${d_2={\rm dn}(Z+Z_0,k)}$.

Double null coordinates $u$ and $v$ can then be introduced by putting 
 $$ T=au+bv, \qquad Z=au-bv, $$ 
 where $a$ and $b$ are positive constants. The metric then takes the standard form for colliding plane waves with aligned linear polarization, namely 
 $$ \d s^2=-2\,e^{-M}\,\d u\,\d v + e^{-U}\Big(e^V\,\d x^2+e^{-V}\,\d y^2\Big), $$ 
 where 
 \begin{eqnarray} 
 e^{-U}&=& {c_1\,c_2\,d_1\,d_2\over(1+ks_1s_2)^2}, \nonumber\\ 
 e^V &=& {1\over(1+k^2)}\, \left({1-ks_2\over1+s_1}\right)^2\,
 {c_1\,d_1\over c_2\,d_2}, \label{metricfns}\\ 
 e^{-M}&=& 2abm^2 {(1+k^2)^3\over(1-k^2)^2} 
 {(1+s_1)^2(1-ks_2)^2\over(1+ks_1s_2)^2}, \nonumber
 \end{eqnarray} 
 in which the arguments of the elliptic functions are either \ ${au+bv-T_0}$ \ or \ ${au-bv+Z_0}$ \ appropriately. It may be noticed that, in the limit in which ${\alpha=0}$ (${k=0}$), $T_0=Z_0=0$ and $2ab=m^{-2}$, these expressions reduce to those of the degenerate Ferrari--Ib\'a\~nez solution \cite{FerIba87b}, which is isomorphic to part of the Schwarzschild space-time inside the horizon.

\section{The colliding plane wave space-time}

The question now is to consider whether or not the above solution in the region in which ${u>0}$ and ${v>0}$ can represent the interaction region of two colliding plane waves. The wavefronts of the two waves in this case can be taken to be the null characteristics ${u=0}$ and ${v=0}$. It is then possible to consider the extension of the solution to prior plane wave and Minkowski regions by simply making the substitutions \ ${u\to u\,\Theta(u)}$ \ and \ ${v\to v\,\Theta(v)}$ \ in the metric functions.

It may first be recalled that the focussing singularity occurs in the interaction region when \ $e^{-U}=0$: \ i.e. in this case when ${\eta=1}$, or ${s_1=1}$, or \ ${c_1={\rm cn}(au+bv-T_0,k)=0}$. \ This corresponds to the horizon of the black hole in the more familiar interpretation of this metric. It is not a curvature singularity. In this context, it is a Cauchy horizon through which the space-time can be extended, but not uniquely.

It is also well known that the metric function $e^{-U}$ satisfies the wave equation, and can therefore be expressed in terms of separate (decreasing) functions of $u$ and~$v$. To demonstrate this explicitly, we introduce the new constants $u_0$ and $v_0$ such that \ ${au_0={1\over2}(T_0-Z_0)}$ \ and \ ${av_0={1\over2}(T_0+Z_0)}$. \ We can then introduce new functions \ ${p=a(u-u_0)}$ \ and \ ${q=b(v-v_0)}$ \ so that \ ${T-T_0=p+q}$ \ and \ ${Z+Z_0=p-q}$. \ In terms of these functions, 
 $$ e^{-U}= {{\rm cn}(p+q)\, {\rm cn}(p-q)\, {\rm dn}(p+q)\, {\rm dn}(p-q)\over \Big(1+k\, {\rm sn}(p+q)\, {\rm sn}(p-q) \Big)^2}, $$ 
 in which the parameter of all elliptic functions is~$k$. Using standard identities, this can be expressed in the form 
 \begin{equation} 
 e^{-U}= 1-\left( \frac{(1+k)\,{\rm sn}\,p}{1+k\,{\rm sn}^2p} \right)^2
 -\left( \frac{(1-k)\,{\rm sn}\,q}{1-k\,{\rm sn}^2q} \right)^2. 
 \label{Upq} 
 \end{equation}

For a vacuum colliding plane wave space-time, it is necessary that the metric function $U(u,v)$ is $C^1$ across the wavefronts. (The remaining metric functions only need to be~$C^0$.) In this case, it can immediately be seen from (\ref{Upq}) that \ ${U_{,u}=0}$ \ when $p=0$, and \ ${U_{,v}=0}$ \ when $q=0$. It is therefore appropriate to take $p=0$ and $q=0$ as defining the wavefronts. This is consistent with labelling the wavefronts as $u=0$ and $v=0$ provided the arbitrary constants are chosen such that $u_0=0$ and $v_0=0$; i.e. ${T_0=0}$ and ${Z_0=0}$. With these choices, the plane of collision between the waves is now identified as the spacelike surface on which ${\eta=k^2}$, and ${\mu=-k}$.

With the plane of collision ${u=0}$, ${v=0}$ now identified, it can be seen that already ${U=0}$ here and hence throughout the background Minkowski region. In order to also have $V=0$ on this plane, it is appropriate to re-scale the $x,y$-coordinates by putting 
 $$ x=\sqrt{1+k^2}\>\tilde x, \qquad y={\tilde y\over\sqrt{1+k^2}}, $$ 
 after which 
 $$ e^V = \left({1-ks_2\over1+s_1}\right)^2\, {{\rm cn}(au+bv)\,{\rm dn}(au+bv)\over 
 {\rm cn}(au-bv)\,{\rm dn}(au-bv)}. $$ 
 It is also appropriate to set $M=0$ on this plane and in the background region by relating the parameters $a$ and $b$ to the mass parameter of the original $C$-metric by putting 
 $$ 2ab ={(1-k^2)^2\over(1+k^2)^3\,m^2}. $$ 
 Thus, in the interaction region 
 $$ e^{-M}= {\Big(1+{\rm sn}(au+bv)\Big)^2\Big(1-k\,{\rm sn}(au-bv)\Big)^2\over
 \Big(1+k\,{\rm sn}(au+bv)\,{\rm sn}(au-bv)\Big)^2}. $$

Throughout the interaction region $u>0$, $v>0$, the non-zero components of the Weyl tensor are given by 
 $$ \Psi_0=-3\,b^2\,e^M\,\Psi(u,v), \qquad \Psi_2=ab\,e^M\,\Psi(u,v), \qquad 
 \Psi_4=-3\,a^2\,e^M\,\Psi(u,v), $$ 
 where 
 $$ \Psi(u,v)= (1-k^2){\Big(1+k\,{\rm sn}(au+bv)\,{\rm sn}(au-bv)\Big)\over
 \Big(1+{\rm sn}(au+bv)\Big)\Big(1-k\,{\rm sn}(au-bv)\Big)}. $$ 
 (This is consistent with this region being of type~D.) However, the above construction implies that impulsive gravitational wave components occur on the boundaries of this region. These are given by the additional components 
 $$ \Psi_0=2b\,e^M\,{{\rm dn}^2(au)-k\,{\rm cn}^2(au) \over 
 {\rm cn}(au)\>{\rm dn}(au)} \>\delta(v), \qquad 
 \Psi_4=2a\,e^M\,{{\rm dn}^2(bv)+k\,{\rm cn}^2(bv) \over 
 {\rm cn}(bv)\>{\rm dn}(bv)}\>\delta(u). $$ 
 These expressions, together with the fact that $\Psi(u,v)$ depends on both \ ${{\rm sn}(au+bv)}$ \ and \ ${{\rm sn}(au-bv)}$, \ indicate an essential asymmetry between the two wave components when~${k\ne0}$.

Using the above procedure to extend the solutions to the prior regions with ${u<0}$ or ${v<0}$, it can be seen that the initial waves before their collision are each a combination of an impulsive and a shock wave. These are given explicitly by 
 $$ \begin{array}{ll} 
 u<0: \qquad &\Psi_0=2b(1-k)\>\delta(v) -3b^2\,\Psi(0,v)\,\Theta(v), \\[6pt]
 v<0: &\Psi_4=2a(1+k)\>\delta(u) -3a^2\,\Psi(u,0)\,\Theta(u). 
 \end{array} $$ 
 The amplitudes of the two waves are determined by the parameters $a$ and~$b$, while the profile of the waves is modified from that of the degenerate Ferrari--Ib\'a\~nez solution by the introduction of the parameter~$\alpha$, which is represented here in~$k$.

\section{Concluding remarks}

It has been demonstrated explicitly above that part of the $C$-metric space-time inside the black hole horizon may be interpreted as the interaction region of colliding plane gravitational waves with aligned linear polarization, provided the rotational coordinate is replaced by a linear one. This is a generalization of the degenerate Ferrari--Ib\'a\~nez solution. It is also an explicit case of the generally asymmetric solutions with a Cauchy horizon given by Feinstein and Ib\'a\~nez~\cite{FeiIba89}, which are expressed in terms of Fourier--Bessel integrals.

An alternative region of this space-time which could have a similar interpretation of this type is obtained by modifying the transformation~(\ref{firsttrans}) by putting ${r=m(1-\eta)}$. This would lead to a different colliding plane wave space-time in which the focussing singularity is a curvature singularity corresponding to that at ${r=0}$ in the $C$-metric.

In view of other generalizations described in \cite{ChaXan86b}--\cite{PanXan89}, it may be conjectured that further colliding plane wave solutions could be obtained by also including rotation, NUT and (electric and magnetic) charge parameters. These may be derived from the complete six-parameter family of type~D solutions given for example by (17)--(19) of~\cite{GriPod06}. It is only required that the interaction region of the colliding plane wave space-time corresponds to a time-dependent region inside the black hole event horizon in the more familiar interpretation. In these solutions, the focussing singularity will generally be a Cauchy horizon rather than a curvature singularity as this would correspond to a black hole horizon. However, it should be emphasised that solutions with a Cauchy horizon are not generally of type~D. As shown in~\cite{FeiIba89}, they are generically algebraically general.

\end{document}